\documentstyle[aps,multicol,prl,epsfig]{revtex}
\begin{document}
\def\sech{\mathop{\rm sech}\nolimits}
\def\csch{\mathop{\rm csch}\nolimits}
\def\coth{\mathop{\rm coth}\nolimits}
\def\span{\mathop{\rm Span}\nolimits}
\def\sn{\mathop{\rm sn}\nolimits}
\def\cn{\mathop{\rm cn}\nolimits}
\def\dn{\mathop{\rm dn}\nolimits}
\def\sign{\mathop{\rm sign}\nolimits}

\begin{title}
{\bf An exploding glass ?}
\end{title}

\author{P.G. Kevrekidis$^1$, S. K. Kumar$^2$ and I.G. Kevrekidis$^3$} 
\address{$^1$ 
Department of Mathematics and Statistics, University of Massachusetts, 
Amherst, MA 01003-4515, USA \\
$^2$ Department of Chemical Engineering, Rensselear Polytechnic Institute, 
Troy, NY 12180  \\
$^3$ Department of Chemical Engineering and PACM, Princeton University,
6 Olden Str. Princeton, NJ 08544 \\}

\date{\today}
\maketitle

\begin{abstract}
We propose a connection between self-similar, focusing dynamics
in nonlinear partial differential equations (PDEs) 
and macroscopic dynamic features of 
the glass transition.
In particular, we explore the divergence of the appropriate
relaxation times in the case of hard spheres as the limit of
random close packing is approached.
%
%
We illustrate the analogy in the critical case, and
suggest a ``normal form'' that can capture the onset of
dynamic self-similarity in both phenomena.

\end{abstract}

\vspace{5mm}

\section{Introduction}

The onset of dynamic self-similarity, {\it i.e.}, the emergence of
solutions that blow-up (or decay) with appropriate scaling laws,
has been a very widely studied topic (see e.g.,
\cite{sulem,Pelinovsky,Rasmussen} and references therein).
Focusing has been extensively examined through a dynamic renormalization
framework.
The key idea lies in appropriately dynamically rescaling  space, time
and the amplitude of the solution \cite{Landman}, so that self-similar blowup solutions
appear as steady states of the renormalized equation.
A benchmark model for blowup is the nonlinear Schr{\"o}dinger (NLS) equation  \cite{sulem},
with which a large number of studies have been concerned \cite{Landman,Wang,Zakharov}.

In a recent study \cite{SKK} we revisited NLS focusing using a template-based, dynamic
renormalization approach motivated by the template-based symmetry 
reduction of  
\cite{RM2000}.
Our procedure can be generally applied in equations that (may) possess   
solutions with dynamic self-similarity. 
Alternative such examples can be found in  \cite{PNAS_never} and
in the more systematic study of \cite{RKML}.
While essentially equivalent to other dynamic renormalization techniques, 
this approach provided a natural setting for studying the 
parametric onset of dynamic self-similarity
{\it as a steady state bifurcation problem} for a PDAE: a PDE (in the 
``co-exploding'' frame) coupled with an algebraic constraint.
%
%
%
The latter imposes the instantaneous rescaling rate by comparing
the evolving solution with rescaled versions of 
an (essentially arbitrary) template function.
At steady state of the PDAE,  one recovers the self-similar shape 
as well as the asymptotic ``explosion rate" $G$.
One also recovers -as with any correct dynamic renormalization approach-   
the scaling rates for time, space, and for various
observables of the system as a function of (original or rescaled) time:
the self-similarity exponents.
The method can be applied to both types of self-similar 
solutions \cite{barenblatt}; it was used in \cite{PNAS_never} to locate 
both the Barenblatt and the Graveleau solutions of the nonlinear diffusion equation. 

The interesting twist is that, 
in this rescaled setting, the appearance of solutions that blow up
(in the NLS setting, in finite time) becomes a steady state 
bifurcation problem: the bifurcation from branches of solutions
with $G=0$ to branches with nonzero $G$. 
The nature of this bifurcation was computationally explored in \cite{SKK},
complementing earlier analytical and numerical studies \cite{sulem,Landman}.
A good analogy to this setting could be given by the bifurcation, 
from a ring of stationary solutions,
of nonzero-speed rotating waves in a pattern formation probem with O(2) 
symmetry \cite{yan}.
%
%
%
%
%
%

The purpose of this note it to propose, at a phenomenological level, a
connection between focusing dynamics of NLS close to criticality, 
on the one hand,
and the onset of macroscopic glassy dynamics  on the other.
Our particular ``glassy example"  is a thought experiment involving
hard-sphere jamming:
it arises from modeling the continuing insertion of hard spheres in a fluid of
such spheres. 
We will show that this non-parametric example exhibits dynamics consistent
with critical blowup (i.e., with the dynamics exactly at the onset of
self-similarity) in NLS.
We will then go one step further in this phenomenological analogy,
and consider the case of a distinguished parameter.
We will speculate about the ``explosion-slowdown" connection 
beyond criticality, motivated by a ``normal form" for the parametric
dependence of the NLS explosion rate.
Beyond criticality, NLS solution observables explode at finite time;
one might conversely think for glasses that time ``blows up" 
at finite values of the system observables.
We will conclude by a brief discussion, and outline 
computational experiments that might confirm or disprove the 
proposed analogy.

\section{A Critical Example: the Sphere Jamming Problem}

As a paradigm of glassy behavior, 
we examine
the evolution of the density of a fluid of hard spheres in the neighborhood 
of the close packing limit. 
%

%
We
are interested in the insertion probability of an additional sphere in
a fluid of hard spheres: 
its  ``materialization" is ``accepted" if
the new sphere does not overlap with any of the previously existing ones
in the fluid.

For the hard sphere system we will consider that our thought 
experiment proceeds at constant 
volume, and the pressure increases, presumably diverging in the limit of
random close packing \cite{torquato2}.

%
%
Let us define the ``density'' (i.e., volume fraction), $\eta$ as
\begin{equation}
\eta=\frac{N}{V} \frac{\pi \sigma^3}{6}
\end{equation}
Here $N$ is the number of spheres, each of diameter $\sigma$ and $V$ is the 
volume of the box. This system undergoes a crystallization 
transition
at $\eta_m\sim$0.495 \cite{torquato,weitz,silescu}. The equation of state 
changes when one crosses this density, even though we shall be 
dealing
with fluid phases at all times. The thermodynamics of these systems are 
governed
by the radial distribution function at contact, i.e., $g(\sigma^+)$. 
For $\eta\le \eta_m$
an excellent approximation is,
\begin{equation}
g(\sigma^+)=\frac{(1-\eta/2)}{(1-\eta)^3}
\end{equation}
This is known as the 
Carnahan-Starling approximation \cite{torquato2,torquato}. For $\eta> \eta_m$, 
Torquato \cite{torquato2} has proposed the
following approximation,
\begin{equation}
g(\sigma^+)=g_m(\sigma^+)\frac{(\eta_c-\eta_m)}{(\eta_c-\eta)}
\end{equation}
where $\eta_c$ is the random close packed density, the ultimate limit for
hard spheres, and $g_m(\sigma^+)$ is the contact value 
of the radial distribution function 
at $\eta_m$. We now use that
\begin{equation}
A(V)-A(V=\infty)=-\int_{\infty}^{V} PdV
\end{equation}
%
%
where $A$ is the Helmholtz energy, $P$ the pressure, and $V$ the volume, 
and further use the identity
\begin{equation}
\frac{PV}{Nk_BT}= 1 + 4\eta g(\sigma^+).
\end{equation}
The probability $p_{ins}$ 
for particle insertion (i.e., taking a system of N particles 
in volume V and
adding one more particle to it) is then computed through
\begin{equation}
-{\rm ln} p_{ins} = A_{{\rm ex}}(N+1) -A_{{\rm ex}}(N),
\end{equation}
where $A_{{\rm ex}}(N)=A(N)-A_{ID}(N)$, and $A_{ID}(N)$ is the Helmholtz
energy of an ideal gas system with $N$ particles in volume $V$.

%
%
Here we begin to perform a thought experiment: we attempt to add speres
to the fluid at a constant attempt rate.
%
%
If this attempt rate is slow compared to the relaxation of the fluid 
after each new addition, 
%
%
we can exploit thermodynamic considerations
(the insertion probability based on the equilibrium radial 
distribution function)
to extract a form of dynamics.
A ``kinetic" evolution of the density of spheres, based in part on our
arbitrary (but slow) attempt rate, and in part on the equilibrated 
(for each density) radial distribution function, would then arise, in the 
form of the ordinary differential equation:

%
%
%
%
%
\begin{eqnarray}
\frac{d \eta}{dt} \sim
\left \{\begin{array}{ll} \exp
\left[-\frac{\eta (8-9 \eta+3 \eta^2)}{(1-\eta)^3}\right]  & \eta \leq 0.495, \\
\exp
\left[-16.543+ 3.576 {\rm ln}\left( \frac{0.644-\eta}{0.644-0.495} \right)
-3.576 \left(\frac{\eta}{0.644-\eta}-\frac{0.495}{0.644-0.495} \right) \right]  &        0.495 \leq \eta \leq 0.644, \\
\end{array}
\right.
\label{add4}
\end{eqnarray}
The right hand side in Eq. (\ref{add4}) represents essentially the
probability of insertion of an additional sphere in the fluid of spheres.
This relation is meant as an equation with a non-dimensional
time (rescaled by 
the ``sphere materialization attempt rate'', $\dot{N} \pi\sigma^3/(6 V)$
[attempts/sec]). We repeat that these
attempts are executed quasistatically, so that equilibrium ideas
can be used to evaluate $p_{ins}$.

We examine briefly the transient behavior induced by the
top line of Eq. (\ref{add4}). The function of the
exponent $f=\eta (8- 9 \eta +3 \eta^2)/(1-\eta)^3$ can be well-approximated
by the Taylor series $f \approx 8 \eta + 15 \eta^2 + 24 \eta^3 + 35
\eta^4 + O(\eta^5)$, while the $\exp(-f)$ is practically indistinguishable
(in the range of values of interest) from $\exp(-\alpha \eta - \beta \eta^2)$,
with $\alpha=8$ and $\beta=15$.
Hence a very accurate approximate solution is given by
\begin{eqnarray}
\eta=\frac{-\alpha+ 2 \sqrt{\beta} 
{\rm Erfi}^{-1} \left[ \frac{2 \sqrt{\beta} (t + C) \exp(\frac{\alpha^2}{4 \beta})}
{ \sqrt{\pi}} \right]}{2 \beta},
\label{add5}
\end{eqnarray}
where $C$ is an appropriate constant chosen by the initial condition
$\eta(t=0)$. Notice that already in this transient dynamics the 
exponential (asymptotic) behavior of the imaginary error function Erfi
(${\rm Erfi}(z) \sim \exp(z^2)/z$, for large z), already
preludes the logarithmic slowdown as higher densities are approached. 
%

We now turn to the case of approach to the 
maximal concentration of 0.644, described by the bottom panel of 
Eq. (\ref{add4}). The latter equation can be cast in the form
\begin{eqnarray}
\frac{ d \eta_r}{d \tau}=- \eta_r^{a_s} \exp\left[-\frac{C_s}{\eta_r} \right],
\label{add6}
\end{eqnarray}
where $\eta_r=6.711 (0.644-\eta)$ is a rescaled measure of the density
(in fact of its ``distance'' from the maximal density); 
$\tau=2.263 t$ is a rescaled time. The constants of the equation
are $a_s=3.576$, and $C_s=15.456$. 

One can straightforwardly identify
Eq. (\ref{add6}) with the normal form equation {\it at criticality}
of a relevant observable (the blowup rate) of the NLS equation which we
will now discuss. In the 
following section, motivated by this analogy, we proceed to give the 
relevant background, examining in more detail the dynamics of focusing in 
the NLS setting.


%
%
%
\section{NLS Background: the Focusing Transition}

\begin{figure}[h]
\centerline{\psfig{file=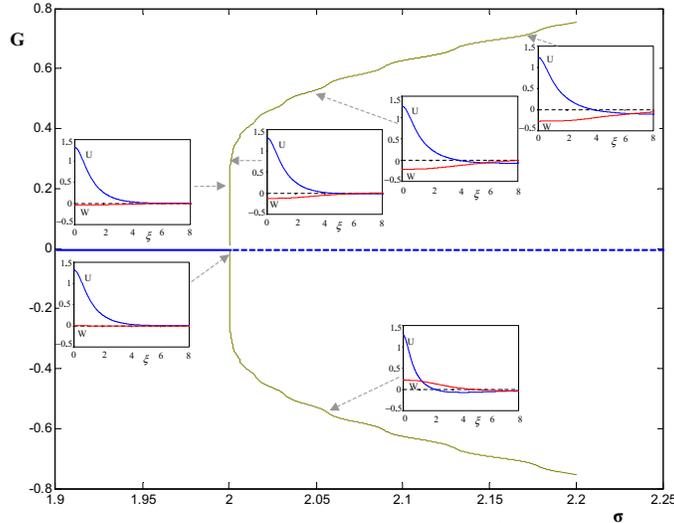,height=3.5in,angle=270}}
 \caption{The figure shows the  bifurcation
diagram of $G$ vs. $\sigma$. At $\sigma_{cr}=2$, the new branch of
focusing solutions is born. The  panel insets show the profile
of the solution in different points along the branch. }
\label{csfig2}
\end{figure}

As mentioned above, the nonlinear Schr{\"o}dinger (NLS) equation \cite{sulem}
is a famous example of an envelope equation
that exhibits a transition from regular to self-similarly blowing up 
solutions.
In the NLS
\begin{eqnarray}
i u_t=-u_{xx} - |u|^{2 \sigma} u,
\label{geq1}
\end{eqnarray}
focusing phenomena (e.g., finite time blowup) are known to occur 
as the parameters (more specifically in the case of Eq. 
(\ref{geq1}), the power $\sigma$ of the nonlinearity) are varied. 
Here $u$ is a complex envelope field (e.g., a slow modulation
of the electric field of light) \cite{sulem}, 
while the subscripts denote partial derivatives. 
Notice that the one
dimensional version of the equation (in space) has been used, but
the problem can also be considered in $d$-dimensions (where the 
second spatial derivative is exchanged with the Laplacian). 
It is
then found \cite{sulem,SKK} that while solitary wave solutions exist
for all values of $\sigma$ in Eq. (\ref{geq1}), for $\sigma \geq \sigma_c
\equiv 2$
(and in general for $d \sigma \geq 2$), such solutions are unstable
towards blowup (collapse). 
In the latter case, the amplitude of the solution 
becomes infinite (in finite time), while the ``width" of the solution
shrinks to 0. 
The two processes occur in a self-similar fashion
(and with rates that are proportional to each other). 
These singular solutions are of the form:
\begin{eqnarray}
u(x,t)=L^{-1/\sigma} e^{i \tau} v(\xi,\tau),
\label{geq2}
\end{eqnarray}
where $\xi=x/L(t)$ is the rescaled spatial variable and $\tau$ given
by $\tau_t=1/L^2$ is the rescaled time. 
$L(t)=2 G (t^{\star}-t)^{1/2}$ is the stretching length, and
$-L_{\tau}/L=G$ is the blowup rate. 
$t^{\star}$ is the (finite)
time at which blowup occurs. 
These solutions can be systematically
analyzed from a dynamic renormalization point of view 
\cite{sulem,Landman,SKK},
in which they appear as (stable) steady states. Using the
scaling of Eq. (\ref{geq2}) or through the systematic
procedure of deriving it \cite{SKK}, one obtains the
renormalized PDE
\begin{eqnarray}
i v_{\tau}=-v_{\xi \xi}-|v|^{2 \sigma} v-
i G(\tau) (\frac{1}{\sigma} v + \xi v_{\xi}).
\label{cseq3}
\end{eqnarray}
An appropriate algebraic constraint is given by
\begin{eqnarray}
\int_{-\infty}^{\infty} Re(v(\xi,\tau)) T(\xi) d \xi= C,
\label{cseq9}
\end{eqnarray}
%
%
where $C$ is an essentially arbitrary constant and
$T(\xi)$ is an arbitrary ``template'' function; this
provides the algebraic equation that determines the 
instantaneous explosion rate
$G(\tau)$. 
Other appropriate formulations of this algebraic constraint (e.g.,
the minimization of the distance of the evolving solution
from scaled versions of the template) are equally acceptable
and yield the right steady state.
Essentially $G$ plays the role of a Lagrange multiplier that enforces 
the conservation of the overlap integral of Eq. (\ref{cseq9}),
according to:
\begin{eqnarray}
G=-\frac{L_{\tau}}{L}= - \frac{\int_{-\infty}^{\infty}
\left[ W_{\xi \xi} + (U^2+W^2)^{\sigma} W - W \right] T(\xi) d \xi}{
\frac{C}{\sigma} + \int_{-\infty}^{\infty} \xi U_{\xi} T(\xi)
d \xi}.
\label{cseq12}
\end{eqnarray}
$U$ represents the real and $W$ the imaginary part of the solution 
$v$ of Eq. (\ref{cseq3}).
Then, the bifurcation of the
focusing branch (with $G \neq 0$) of solutions from the (unstable
for $\sigma > 2$) solitary wave one (with $G=0$) occurs
as a supercritical Hopf-like bifurcation in a new type (mixed Hamiltonian-
dissipative; see e.g., \cite{SKK}) of dynamical system. 
In the relevant bifurcation diagram shown in Fig. \ref{csfig2}, 
the appearance of the new (focusing) 
branch occurs, as the parameter $\sigma$ is
varied, with an exponentially small dependence
according to \cite{sulem,Landman,SKK}
\begin{eqnarray}
 \epsilon G^{ss} \sim \exp(-\frac{C}{G^{ss}}),
\label{geq3}
\end{eqnarray}
where $\epsilon= (\sigma-2)/\sigma$ and the superscript ``ss'' serves
to denote that this equation yields the steady state value of $G$
(i.e., the parameter-dependent, asymptotic blowup rate).
We can think of G as an observable of the transient solution of the
PDAE set of Eqs. (\ref{cseq3})-(\ref{cseq12});
it is the rate by which the width of the solution must be scaled 
(and proportional to
the rate by which the amplitude of the solution must be scaled), so
that the solution appears as steady.
%
%
We will return later to the parametric dependence of the asymptotic blowup
rate as a function of the parameter $\epsilon$, and even the dynamical 
approach
of $G$ to its steady state value in transient simulations.

We will now concentrate on the critical case $\sigma=2$, where the 
asymptotic $G$ 
is zero; indeed, at criticality, 
all the (appropriately) rescaled forms of the soliton are standing wave
solutions of the NLS; hence, the
soliton is neutral to scale perturbations.
%
%
%
In this critical case, it is known from earlier work \cite{sulem,Landman}
that there is no solution 
with $G \neq 0$; if one starts close to (but not exactly on) a solution 
with $G=0$, the
evolution of $G$ will follow the normal form dynamics \cite{sulem,Landman}
of
\begin{eqnarray}
\frac{d G}{d \tau} = - G^{a} \exp\left( -\frac{C}{G} \right),
\label{add1}
\end{eqnarray}
where $a=-1$, and $C \approx 3.1$.

The steady state branch results in \cite{sulem,Landman,SKK} 
and the dynamic results for the critical case normal form of Eq. (\ref{add1}) 
can be unified \cite{skk2} in a single ``normal form'' equation 
containing the salient features of the onset of dynamic self-similarity for
the NLS:
namely
\begin{eqnarray}
\frac{d G}{d \tau} \sim \epsilon - G^{a} \exp\left( -\frac{C}{G} \right).
\label{add2}
\end{eqnarray}
%
%
%

This equation encapsulates the slow (logarithmic) approach to
$G=0$  for the critical case of $\sigma=2 (\epsilon=0)$.
Setting $dG/d \tau =0$ at supercritical $\sigma$, the
equation will give the correct 
stable blowup solution. 
%
%
As a dynamic equation with an equal sign, for $\sigma > 2$ it will predict 
the ``relaxation'' (in fact, reshaping) of the now unstable soliton 
into the stable blowup solution, and
the subsequent asymptotic evolution of the latter towards infinite 
amplitude. 
The details of this normal form for the
onset of dynamic self-similarity will be presented elsewhere \cite{skk2}.
We now turn to the connections of this normal form at criticality
with our particular glassy example.

\section{From  Focusing to Jamming: Drawing Analogies}

It is interesting to draw the parallels between Eq. (\ref{add6}) and Eq. (\ref{add1}),
and between the two cases more generally.
In our
example of glassy behavior (the sphere problem), there is no free 
parameter.
Hence, it is natural to expect that this example is exactly on the threshold
(critical point) for the onset of glassy behavior (i.e., at ``$T=T_g$'', or
equivalently at $\sigma=2$). Naturally then one expects that 
the concentration approaches the close packing limit, but with a 
logarithmic (``glassy'') slowdown. The system ``arrives'' at such a 
limiting  density in infinite time (i.e., the time explodes at finite
density) but only at an extremely slow rate, as is the case for
the weak collapse of NLS in the critical setting of $\sigma=2$.

Should a natural 
parameter
be inserted in the system, we would expect that its behavior 
would exhibit a transition to 
dynamically self-similar ``glassiness'' as a critical parameter value is crossed.
We expect that the more general NLS ``normal form" dynamics could 
account for the glass transition as such a bifurcation: another onset of 
dynamic
self-similarity.
This time, however, instead of explosive acceleration, a progressive slowdown
of the solution observables/rates is expected.

Notice that one can further deduce, using appropriate integral
asymptotic expansions \cite{peli}, the asymptotics of the approach
to the maximal density of 0.644.
Using the transformations 
$\tau \rightarrow \tilde{\tau}=1.156 \times 10^{3} \tau$ and
$\eta_r \rightarrow \tilde{\eta}= 15.646/\eta_r$, one obtains
the asymptotic expression for the time dependence of the density
\begin{eqnarray}
\tilde{\eta} \sim {\rm ln}(\tilde{\tau}-\tilde{\tau}_0)
\left[1- 1.576 \frac{{\rm ln}
\left( {\rm ln}(\tilde{\tau}-\tilde{\tau}_0) \right)}
{{\rm ln}(\tilde{\tau}-\tilde{\tau}_0)} \right].
\label{add7}
\end{eqnarray}

\begin{figure}[h]
\centerline{\psfig{file=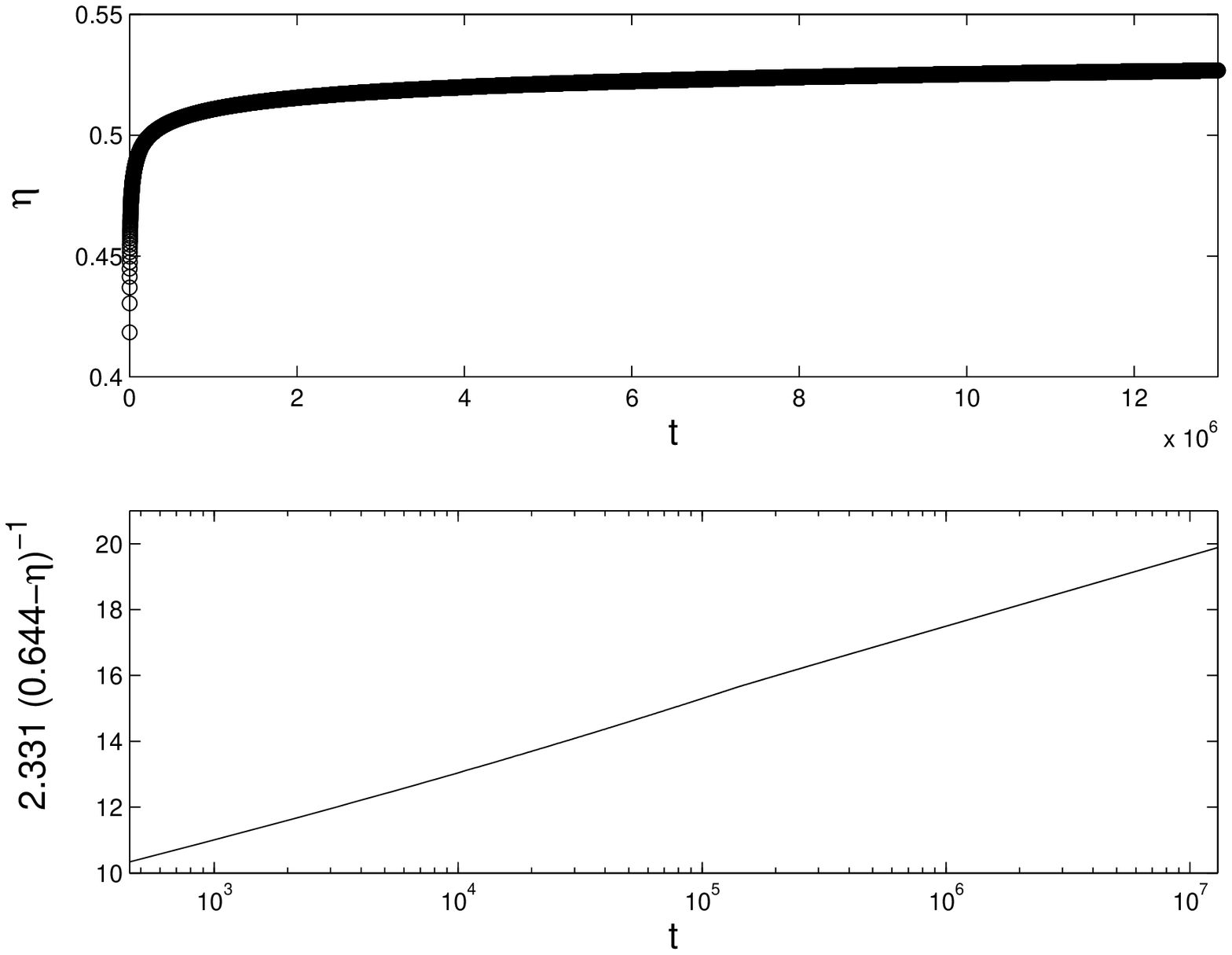,height=3.5in,angle=0}}
 \caption{The top panel shows the evolution of the density
$\eta$ as a function of time, from an initial condition of
$\eta(0)=0.1$. The bottom panel illustrates in a semilog plot
the logarithmic nature of this dependence in plotting $\tilde{\eta}$
as a function of $t$. The linear best fit is yields a slope of
$0.95$ which is very close to the theoretically predicted slope
of $1$. The discrepancy can be well-justified by the asymptotic
nature of the prediction as well as by the doubly logarithmic
corrections of Eq. (\ref{add7}).}
\label{skfig2}
\end{figure}

A very long time evolution (for times of $O(10^6)$) of the ODE
of Eq. (\ref{add4}) is shown in Fig. \ref{skfig2}, verifying
the prediction for logarithmic slowdown as the close packing
density is approached.
%
%
One might argue that, what is referred to as ``the colloidal glass transition"
occurs when the observed rate becomes, say, twelve orders of magnitude
slower than the initial rate.
Using this argument, one would ``declare" a colloidal glass 
transition from this diagram at a density of 
$\eta \approx 0.566$, within the accepted 
range of $0.56-0.58$ \cite{weitz}. 
Notice, however, that in experimental
studies of the colloidal glass transition, the particles are moving
even for higher concentrations, and they only freeze (and the viscosity 
diverges) at a density of 0.644 \cite{chaik}.
%
%

Finally, we also mention in passing that using the results of
\cite{torquato2}, one can infer the leading order spatial
and temporal dependence of the conditional pair distribution function.
Its amplitude will grow as $\sim 1/(\eta_c-\eta)$ i.e.,
logarithmically in time, while it will decay to leading order
as $1/r$ in space (for large $r$). We will examine in more
detail the spatial dependence of properties of the glass
transition in a future publication.

\section{Discussion}

The description of the glassy state has been a puzzling problem
for theorists for many decades \cite{pgd,fhs,rev2}. 
On the one hand, the transition to this state
does not possess as clean-cut characteristics
as other phase transitions \cite{pgd};
on the other hand, however, it is clear
that vitrification induces a significant change in the properties
of a material. 
Furthermore, often but not always a critical
(glass transition) temperature, $T_g$, can be identified \cite{gotze}.

For these reasons, the glass transition is often considered/studied
as a {\it dynamical transition}. 
The relevant prism that is at present widely
used \cite{pgd,fhs} is that of mode coupling theory 
\cite{leuth} that views the formation of the glass as a transition
from ergodic to non-ergodic behavior in which structural arrest
renders whole regions of phase space unavailable for microscopic
configurations. 
In this framework, a macroscopic correlation function
(in fact, the Fourier transform of the density-density
correlation function) is the central object of study, and linear
(or nonlinear) ordinary differential equations with delay
are written for its time evolution. 
The existence of memory kernels in the time
evolution results in the evolution towards
a different ``steady state'' for long time
asymptotics in the glassy state than in the liquid case. 
This
dynamical transition is identified as the glass transition \cite{pgd}.

Motivated by the previous sections, we posit an alternative viewpoint
on this transition.
We showed an analogy between the logarithmic dynamics of the NLS at
criticality, and the logarithmic behavior of the density in the
sphere jamming problem.
Based on this analogy, we argue that the non-parametric sphere-jamming
problem is ``critical" in its glassy dynamics.
%
%
An important link in our analogy is the use (in our thought experiment)
of ``thermodynamic
information'' (an equilibrated radial distribution function) with a 
kinetic concept (a constant sphere addition attempt rate) to
predict ``glassy" dynamics.
We conjecture that, when an equation corresponding to Eq. (\ref{add2})
becomes available,
containing a distinguished parameter, 
dynamics analogous to the supercritical focusing NLS will emerge.
%
%
While for NLS the rates of change of some observables accelerate with time 
(and the corresponding observables ``explode" to infinity), 
for glasses the analogous rates will decelerate with time.
When observables like density approach a finite value, the inverse of the
distance of the observable from its limiting value again approaches infinity
(see Eq. (\ref{add7})); 
zeros and infinities in the two cases appear to be a matter
of choice of observables.

Typical dynamic codimension-one bifurcations result in branches that are power-laws
in the bifurcation parameter (see e.g., \cite{GuckHolmes}).
In contrast, in the NLS case, the onset of dynamic self-similarity is an
exponentially small effect.
We expect a similar structure in the ``onset of glassiness" as a function of
the distance from the corresponding critical point.
%
One can contrast the explosion of the NLS amplitude as a function of time,
to ``time explosion" for the glassy case as a function of an observable 
such as density.

Even though qualitative, and of purely macroscopic nature, the 
conjecture carries a number of 
attractive  features as an alternative viewpoint on the glass transition.
A natural next step would involve the examination of the sphere
model in the presence of an external tunable free parameter.
%
Models of glassy dynamics possessing a distinguished parameter might
also provide supporting evidence \cite{chandler}.
%
%
In \cite{SKK} we computationally converged on self-similar
solutions through either integration or bifurcation analysis of the
dynamically renormalized equation.
It is possible to extend this procedure to operate on
a microscopic (e.g. molecular dynamics) simulator of
a process, when we believe that macroscopic equations of the type of 
Eqs. (\ref{cseq3})-(\ref{cseq12})
exist, but are not available in closed form (e.g. see \cite{manifesto}).
In our case we were fortunate to have a single closed equation in terms
of a single observable (the density); in more complicated systems the
``coarse dynamics'' approach can be used to search for self-similar dynamics
in the evolution of several moments of an evolving particle distribution
without ever writing these equations in closed form.
It is conceivable that a distributed macroscopic observable (like the 
density-density correlation function) can be used 
as the object for which self-similar dynamics in space as well
as in time will be sought.
Another interesting direction would be to explore possible analogies
between multiple interacting blowup spots in the focusing/NLS setting 
and the effect of heterogeneities in glassy dynamics of extended systems.
Finally, it would also be relevant to examine
how this view of the glass transition could be related to the
explosion-collapse duality arising in some focusing
problems such as Bose-Einstein condensation \cite{ghosh}.


Acknowledgements

This work was partially supported by AFOSR and NSF/ITR (I.G.K.), as
well as NSF (DMS-0204585), the Clay Mathematics Institute and the 
University of Massachusetts (P.G.K.). We are grateful to P. Chaikin,
W.B. Russel and especially P.G. Debenedetti for discussions.

\end{document}